\documentclass[prl,longbibliography,twocolumn,superscriptaddress,amsmath,amssymb,verbatim]{revtex4-1}

\usepackage{graphicx}
\usepackage{lmodern}
\usepackage{epstopdf}
\usepackage{dcolumn}
\usepackage{bm}
\usepackage{subfigure}
\usepackage{amsmath} 

\usepackage[pdftex,colorlinks=true,citecolor=blue,linkcolor=blue,urlcolor=blue,bookmarks=true]{hyperref}
\usepackage[sort&compress]{natbib}

\usepackage{color}
\usepackage{textcomp}
\definecolor{red}{rgb}{1,0,0}

\definecolor{blue}{rgb}{0,0,1}

\definecolor{green}{rgb}{0,1,0}

\begin{document}
\preprint{APS}

\title{Elementary excitation in the spin-stripe phase in quantum chains}

\author{Matej Pregelj}
\email{matej.pregelj@ijs.si}
\affiliation{Jo\v{z}ef Stefan Institute, Jamova 39, 1000 Ljubljana, Slovenia}
\author{Andrej Zorko}
\affiliation{Jo\v{z}ef Stefan Institute, Jamova 39, 1000 Ljubljana, Slovenia}
\author{Matja\v{z} Gomil\v{s}ek}
\affiliation{Jo\v{z}ef Stefan Institute, Jamova 39, 1000 Ljubljana, Slovenia}
\affiliation{Centre for Materials Physics, Durham University, South Road, Durham, DH1 3LE, UK}
\author{Martin Klanj\v{s}ek}
\affiliation{Jo\v{z}ef Stefan Institute, Jamova 39, 1000 Ljubljana, Slovenia}
\author{Oksana Zaharko}
\affiliation{Laboratory for Neutron Scattering, PSI, CH-5232 Villigen, Switzerland}
\author{Jonathan S. White}
\affiliation{Laboratory for Neutron Scattering, PSI, CH-5232 Villigen, Switzerland}
\author{Hubertus Luetkens}
\affiliation{Laboratory for Muon Spin Spectroscopy, PSI, CH-5232 Villigen, Switzerland}
\author{Fiona Coomer}
\affiliation{ISIS Facility, Rutherford Appleton Laboratory, Chilton, Didcot, Oxon OX11 OQX, 
United Kingdom}
\author{Tomislav Ivek}
\affiliation{Institute of Physics, Bijeni\v{c}ka c. 46, HR-10000 Zagreb, Croatia}
\author{David Rivas G\'{o}ngora}
\affiliation{Institute of Physics, Bijeni\v{c}ka c. 46, HR-10000 Zagreb, Croatia}
\author{Helmuth Berger}
\affiliation{Ecole polytechnique f\'{e}d\'{e}rale de Lausanne, CH-1015 Lausanne, Switzerland}
\author{Denis Ar\v{c}on}
\affiliation{Jo\v{z}ef Stefan Institute, Jamova 39, 1000 Ljubljana, Slovenia}
\affiliation{Faculty of Mathematics and Physics, University of Ljubljana, Jadranska c. 19, 1000 Ljubljana, Slovenia}

\date{\today}

\begin{abstract}

Elementary excitations in condensed matter capture the complex many-body dynamics of interacting basic entities in a simple quasiparticle picture.
In magnetic systems the most established quasiparticles are magnons, collective excitations that reside in ordered spin structures, and spinons, their fractional counterparts that emerge in disordered, yet correlated spin states.
Here we report on the discovery of elementary excitation inherent to spin-stripe order that represents a bound state of two phason quasiparticles, resulting in a wiggling-like motion of the magnetic moments. 
We observe these excitations, which we dub ``wigglons'', in the frustrated zigzag spin-1/2 chain compound $\beta$-TeVO$_4$, where they give rise to unusual low-frequency spin dynamics in the spin-stripe phase.
This provides insights into the stripe physics of strongly-correlated electron systems.

\end{abstract}

\maketitle

\section{Introduction}

The concept of elementary excitations provides an elegant description of dynamical processes in condensed matter \cite{nakajima1980physics}.
Its use is widespread and represents the theoretical foundation for our understanding of vibrational motions of atoms in crystals as phonons \cite{nakajima1980physics}, the excitations of the valence electrons in metals as plasmons \cite{pines2016emergent}, the bound states of an electron and an electron hole in semiconductors as excitons \cite{frenkel1931transformation}, etc.
In magnetic systems, this approach inspired the spinon picture of fractional excitations in spin liquids \cite{balents2010spin}, the phason description of the modulation-phase oscillations in amplitude modulated structures \cite{blanco2013phasons}, and the magnon picture of collective spin excitations in ordered states \cite{bloch1930theorie}.
The latter led to further intriguing discoveries, including longitudinal Higgs modes in two-dimensional antiferromagnets \cite{jain2017higgs} and magnon bound states in ferromagnetic spin-1/2 chains \cite{torrance1969excitation}. 
Yet, for systems where several order parameters interact, the elementary excitations remain mysterious.
A prominent example are the elusive excitations that cause the melting of charge-stripe order in high-temperature superconductors \cite{parker2010fluctuating, fernandes2014drives, fradkin2015colloquium, wang2016strong, klauss2012musr, kivelson2003detect, vojta2009lattice} and promote enigmatic charge fluctuating-stripe (nematic) states \cite{kivelson2003detect, vojta2009lattice, anissimova2014direct}.

\begin{figure}[!]
\centering
\includegraphics[width=\columnwidth]{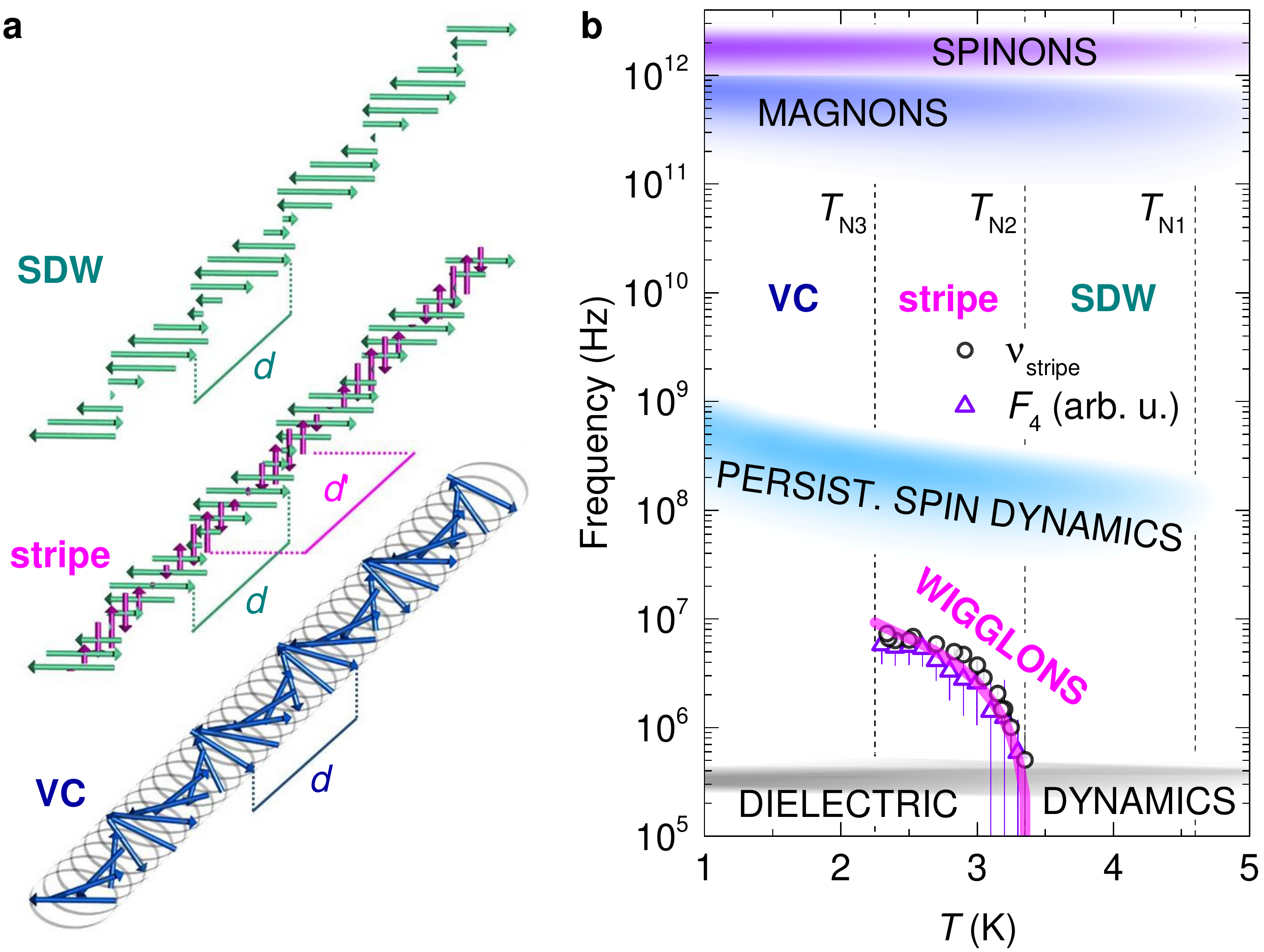}
\caption{{\bf Magnetic structures and magnetic phase diagram.} {\bf a} Magnetic structure models corresponding to the spin-density-wave (SDW), spin-stripe and vectro chiral (VC) phases in $\beta$-TeVO$_4$. Note the difference in modulation periods of the two ordered components in the spin-stripe phase, $d$ and $d'$, associated with the magnetic wave vectors {\bf k} and {\bf k}+{\bf$\Delta$k}, respectively.
{\bf b} The corresponding excitations extend over a wide frequency range, from low-energy dielectric dynamics up to high-energy spinon excitations, i.e., from $\sim$10$^5$\,Hz up to $\sim$10$^{12}$\,Hz.
The derived frequency of the magnetic-order dynamics $\nu_{\text{stripe}}$ in the spin-stripe phase corresponds well to the product of intensities of the two neutron reflections, associated with the fourth-order coupling term $F_4$, and to the phenomenological model (solid line).
The error bars represent an uncertainty of 1 s.d.
}
\label{fig-schem}
\end{figure}

Here we study elementary excitations of the spin-stripe phase in the frustrated spin-1/2 chain compound $\beta$-TeVO$_4$ \cite{savina2011magnetic, gnezdilov2012low, pregelj2015spin, savina2015study, weickert2016magnetic, pregelj2016exchange, pregelj2018coexisting}, which contains localized V$^{4+}$ ($S$\,=\,1/2) magnetic moments \cite{pregelj2015spin, pregelj2016exchange}.
This intriguing order involves two superimposed orthogonal incommensurate amplitude-modulated magnetic components with slightly different modulation periods (Fig.\,\ref{fig-schem}a), corresponding to two magnetic order parameters, that result in a nanometer-scale spin-stripe modulation \cite{pregelj2015spin}. 
We show that in the low-frequency (megahertz) range this, otherwise long-range-ordered state, is in fact dynamical due to the presence of a low-energy excitation mode that results from the binding of two phasons from the two orthogonal magnetic components.
This type of elementary excitation, which we dub a ``wigglon'', is inherent to the spin-stripe order (Fig.\,\ref{fig-schem}) and provides insights into the dynamics of stripe phases that may be found when there are two or more order parameters coupled together.

\begin{figure*}[!]
\centering
\includegraphics[width=17cm]{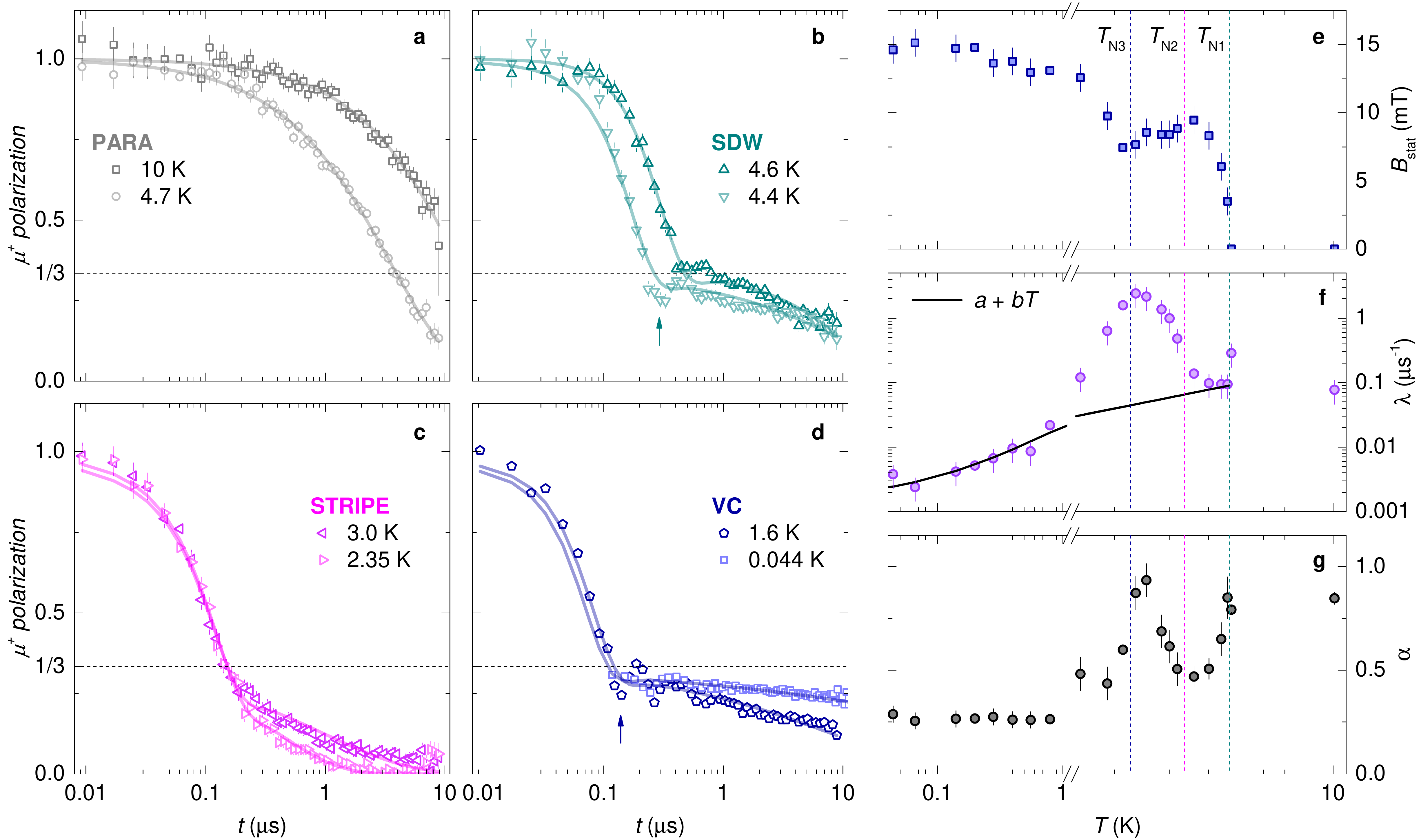}
\caption{{\bf $\mu$SR results.} {\bf a}-{\bf d} The $\mu$SR polarization data. Lines are fits to the model Eq. (\ref{muPolSimple}). The arrows indicate the first minimum due to oscillations caused by static local magnetic fields at 4.4\,K and 1.6\,K. 
{\bf e} The derived mean value of the static local magnetic fields at the muon stopping site.
{\bf f} The derived relaxation rate associated with the local magnetic-field fluctuations. 
The line is a fit to a simple linear model $\lambda\,\propto\,a+bT$. 
The apparent gap in the model at 1 K is just a gap in the horizontal scale.
{\bf g} The derived stretching exponent, reflecting the distribution of the local magnetic-field fluctuations.
The error bars represent an uncertainty of 1 s.d.
}
\label{fig-musr}
\end{figure*}

The peculiar spin-stripe order in $\beta$-TeVO$_4$ evolves from a spin-density-wave (SDW) phase, which
develops below $T_{N1}$\,=\,4.65\,K and  is characterized by a single collinear incommensurate amplitude-modulated magnetic component (Fig.\,\ref{fig-schem}a). 
On cooling, a second superimposed incommensurate amplitude-modulated component with a different modulation period and orthogonal polarization emerges (Fig.\,\ref{fig-schem}a) at $T_{N2}$\,=\,3.28\,K and the spin-stripe order is formed.
Finally, at $T_{N3}$\,=\,2.28\,K the modulation periods of the two incommensurate amplitude-modulated components become equal and a vector-chiral (VC) phase (Fig.\,\ref{fig-schem}a) is established.
For frustrated spin-1/2 chains with ferromagnetic nearest-neighbor and antiferromagnetic next-nearest-neighbor exchange interactions, which in $\beta$-TeVO$_4$ amount to $J_1$\,$\approx$\,$-$38\,K and $J_2$\,$\approx$\,$-$0.8\,$J_1$, respectively \cite{pregelj2015spin}, the SDW and VC phases are predicted theoretically \cite{sudan2009emergent, hikihara2008vector}, while the intermediate spin-stripe phase is not.
The formation of the latter has been associated with exchange anisotropies and interchain interactions \cite{pregelj2015spin,pregelj2016exchange}, but still awaits a comprehensive explanation.

\section{Results}

We explore the spin dynamics in these phases by employing the local-probe muon-spin-relaxation ($\mu$SR) technique, which is extremely sensitive to internal magnetic fields and can distinguish between fluctuating and static magnetism in a broad frequency range (from $\sim$100\,kHz to $\sim$100\,GHz) \cite{yaouanc2011muon}.
We used a powder sample obtained by grinding single crystals (see Methods) to ensure that on average 1/3 of the muon polarization was parallel to the local magnetic field.
In the case of static local fields, $B_{\text{stat}}$, the corresponding 1/3 of the total muon polarization is constant, resulting in the so-called ``1/3-tail'' at late times in the $\mu$SR signal  \cite{yaouanc2011muon}. 
The remaining muon polarization precesses with the angular frequency $\gamma_\mu B_{\text{stat}}$ ($\gamma_\mu$\,=\,2$\pi\times$135.5\,MHz/T), leading to oscillations in the time dependence of the $\mu$SR signal around the ``1/3-tail'' \cite{yaouanc2011muon, sup}.
The only way for the muon polarization to relax below 1/3 at late times is thus provided by dynamical local fields.

At $T$\,=\,4.7\,K\,$>$\,$T_{N1}$, the measured $\mu$SR polarization decays monotonically (Fig.\,\ref{fig-musr}a), as expected in the paramagnetic state where fluctuations of the local magnetic fields are fast compared to the muon lifetime \cite{yaouanc2011muon}.
The muon relaxation curve changes dramatically at $T_{N1}$ (Fig.\,\ref{fig-musr}b), where the polarization at early times suddenly drops, reflecting the establishment of static internal fields in the SDW phase. 
The corresponding oscillations are severely damped, i.e., only the first oscillation at $t$\,$<$\,1\,$\mu$s can be clearly resolved (Fig.\,\ref{fig-musr}b), which indicates a wide distribution of $B_{\text{stat}}$, a hallmark of the incommensurate amplitude-modulated magnetic order.
Clearly, in $\beta$-TeVO$_4$, this static damping is sufficiently strong that the $\mu$SR signal beyond $\sim$1\,$\mu$s can be attributed solely to the ``1/3-tail''.
The latter notably decays (Fig.\,\ref{fig-musr}b), which proves that the local magnetic field is still fluctuating, as expected for incommensurate amplitude-modulated magnetic structures \cite{pregelj2012persistent}.
Remarkably, below $T_{N2}$, in the spin-stripe phase, the ``1/3-tail'' is dramatically suppressed and the oscillation is lost (Fig.\,\ref{fig-musr}c), revealing a significant enhancement of local-field fluctuations.
This indicates that the system enters an intriguing state that is completely dynamical on the $\mu$SR timescale.
Finally, below $T_{N3}$ the slowly-relaxing ``1/3-tail'' and the oscillation reappear (Fig.\,\ref{fig-musr}d), corroborating the establishment of a quasi-static VC state with almost fully developed magnetic moments.

To quantitatively account for the $\mu$SR signal we model the $\mu$SR polarization over the whole temperature range as a product of the two factors 
\begin{align}
\begin{split}
\label{muPolSimple}
P(t)  =  \left[\frac{1}{3} + \frac{2}{3} \cos(\gamma_\mu B_{\text{stat}} t)e^{-(\gamma_\mu \Delta t)^2/2}\right] e^{-(\lambda t)^\alpha}.  
\end{split}
\end{align}
The exponential in the first factor in (\ref{muPolSimple}) accounts for the muon relaxation due to a Gaussian distribution of static magnetic fields with a mean value $B_{\text{stat}}$ and a width $\Delta$.
Since oscillations of the $\mu$SR polarization are almost completely damped already after the first visible minimum, the parameters $B_{\text{stat}}$ and $\Delta$ must be comparable.
Indeed, the best agreement with experiment was achieved for $\Delta/B_{\text{stat}}$\,=\,1.25(1) (Fig.\,\ref{fig-musr}a-d), which was kept fixed for all temperatures.
The second factor in (\ref{muPolSimple}) is the stretched-exponential function that describes the decay of the ``1/3-tail'' due to additional local magnetic-field fluctuations.
Here, $\lambda$ is the mean relaxation rate while $\alpha$ is the stretching exponent accounting for a distribution of relaxation rates \cite{johnston2006stretched}.

The results of our fits of the $\mu$SR data to Eq.\,(\ref{muPolSimple}) are summarized in Fig.\,\ref{fig-musr}.
$B_{\text{stat}}$ (Fig.\,\ref{fig-musr}e) grows from zero at $T_{N1}$ to 9(1)\,mT at $T_{N2}$, which is a value of dipolar fields typical encountered by muons in spin-1/2 systems \cite{yaouanc2011muon}.
In the spin-stripe phase, $B_{\text{stat}}$ slightly decreases, while below $T_{N3}$ it starts growing again and reaches a 15(1)\,mT plateau at the lowest temperatures.
On the contrary, the relaxation rate $\lambda$ does not change significantly throughout the SDW phase, but it escalates by more than an order of magnitude below $T_{N2}$, i.e., in the spin-stripe phase.
Below $T_{N3}$, however, it reduces and resumes following the same linear temperature dependence (solid lines in Fig.\,\ref{fig-musr}f) as in the SDW phase.
This is a characteristic of the persistent spin dynamics \cite{de2006spin} of the disordered part of the magnetic moments in amplitude-modulated magnetic structures \cite{pregelj2012persistent}.
As these fluctuations are fast compared to the muon precession, i.e., they do not suppress the minimum in the $\mu$SR signal described by the first factor in Eq.\,(\ref{muPolSimple}), one can assume that $\lambda$\,=\,2$\gamma_\mu^2 B_{\text{dyn}}^2$/$\nu_{\text{dyn}}$ \cite{yaouanc2011muon}, where $B_{\text{dyn}}$ is the size of the fluctuating field and $\nu_{\text{dyn}}$ is the corresponding frequency.
Considering that in amplitude-modulated magnetic structures $B_{\text{dyn}}$ is comparable to $B_{\text{stat}}$, we can estimate that $\nu_{\text{dyn}}$ ranges between 0.1 and 1\,GHz (Fig.\,\ref{fig-schem}b).
Finally, the stretching exponent $\alpha$ in the SDW and VC phase (Fig.\,\ref{fig-musr}g) amounts to 0.43(5) and 0.25(2), respectively, as expected for broad fluctuating-field distributions in the incommensurate amplitude-modulated magnetic structures \cite{pregelj2012persistent}.

While the $\mu$SR response in the SDW and VC phases is within expectations, the spin-stripe phase shows a surprising enhancement of $\lambda$ (Fig.\,\ref{fig-musr}f) and $\alpha$ (Fig.\,\ref{fig-musr}g) that reflects the severe decay of the ``1/3-tail'' in this phase (Fig.\,\ref{fig-musr}c). 
This clearly demonstrates the appearance of an additional relaxation channel that is related to the spin-stripe order only.
Moreover, the increase of $\lambda$ is accompanied with the loss of the oscillation in the $\mu$SR signal, which indicates that the corresponding fluctuations are associated with the ordered part of the magnetic moments.
To account for these experimental findings we introduce the dynamics of the magnetic order into our minimal model of Eq.\,(\ref{muPolSimple}) via the strong collision approach \cite{yaouanc2011muon}.
Namely, we assume that in the spin-stripe phase the static fields derived for the SDW phase fluctuate with a single correlation time 1/$\nu_{\text{stripe}}$, where $\nu_{\text{stripe}}$ is the fluctuating frequency, and numerically calculate the resulting muon polarization function in a self-consistent manner.
Indeed, the resulting muon polarization function $P_{\text{stripe}}(t)$ \cite{sup}, with all other parameters fixed to the values derived for the SDW phase, explains the response of the $\mu$SR signal throughout the spin-stripe phase \cite{sup}.
The derived temperature dependence of $\nu_{\text{stripe}}$ exhibits a continuous increase from 0.5(5)\,MHz at $T_{N2}$ to 7.3(5)\,MHz at $T_{N3}$ (Fig.\,\ref{fig-schem}b).


%
\begin{figure*}[!]
\centering
\includegraphics[width=15.5cm]{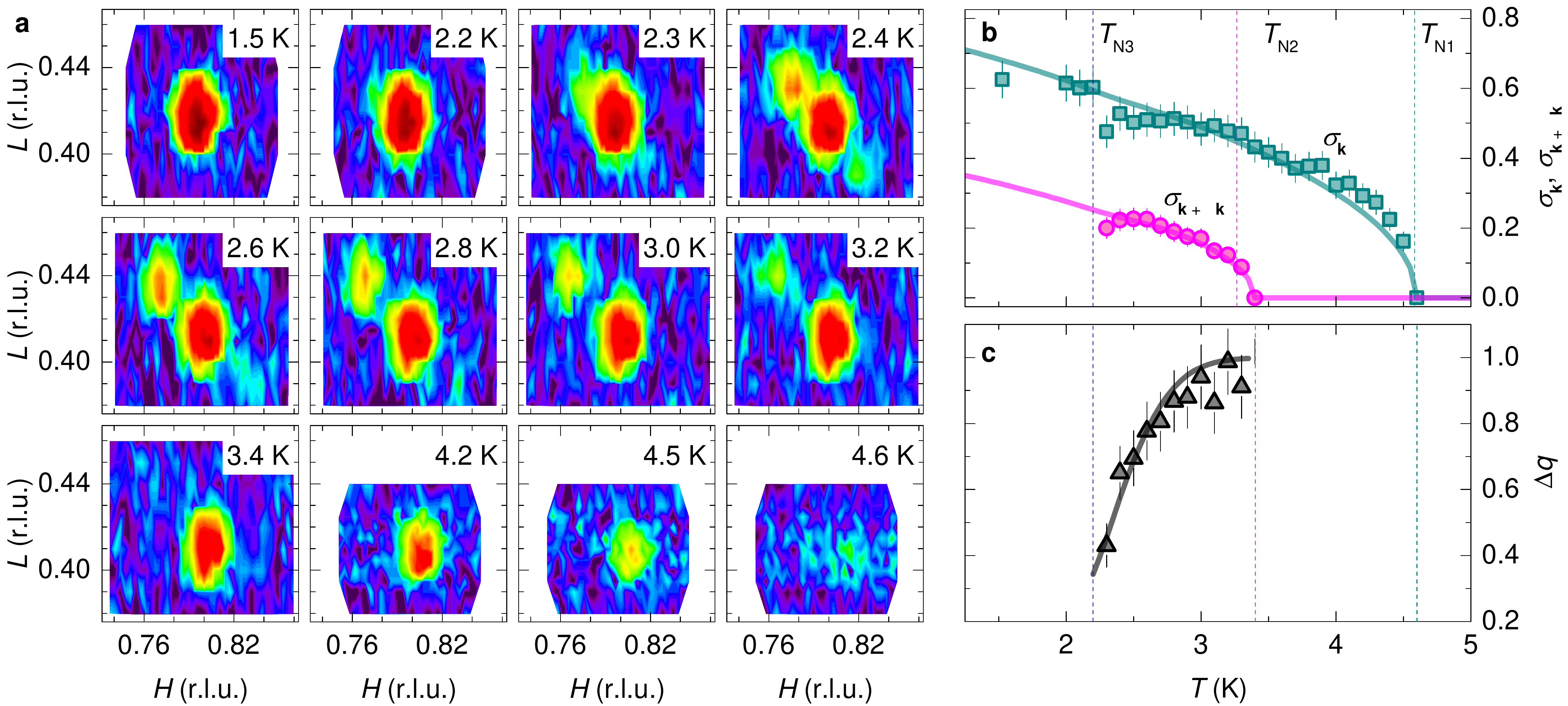}
\caption{{\bf Neutron diffraction results.} {\bf a} The temperature dependence of the strongest magnetic reflection and the accompanying satellite reflections.
The temperature dependences of {\bf b} the order parameters $\sigma_{\bf k}$ and $\sigma_{{\bf k+\Delta k}}$ as well as {\bf c} $\Delta q$ derived from the neutron diffraction experiment (symbols) and calculated using a phenomenological model (solid lines; see text). 
The error bars represent an uncertainty of 1 s.d.
}
\label{fig-NDphen}
\end{figure*}

To further investigate the relation between the spin-stripe order  witnessed previously by neutron diffraction \cite{pregelj2015spin, pregelj2016exchange} and the stripe dynamics observed by $\mu$SR, we performed additional neutron diffraction measurements (see Methods). 
We measured the temperature dependence of the strongest magnetic reflection and its satellites (Fig.\,\ref{fig-NDphen}a), the latter being associated with the orthogonal $b$ magnetic-moment component (Fig.\,\ref{fig-schem}a) \cite{pregelj2016exchange} that emerges at slightly different wave vectors, shifted by $\pm${\bf $\Delta$k} from the main magnetic wave vector {\bf k} \cite{pregelj2015spin}.
The intensity of an individual magnetic reflection, scales with the square of the corresponding order parameter $\sigma_{\bf k}$\,=\,$M_{\bf k}/\mu_B$, where $M_{\bf k}$ denotes the sublattice magnetization component associated with {\bf k} and $\mu_B$ is the Bohr magneton \cite{pregelj2016exchange}.
This allows for the comparison of $\nu_{\text{stripe}}(T)$ with the temperature evolution of $F_4$\,$\sim$\,$\sigma_{\bf -k}^2 \sigma_{\bf k+\Delta k}\sigma_{\bf k - \Delta k}$, i.e., the lowest-order term in the magnetic free energy that couples all magnetic components with different modulation periods ({\bf k}, {\bf k+$\Delta$k}, and {\bf k$-\Delta$k}) \cite{pregelj2016exchange}.
We find a very good correspondence (Fig.\,\ref{fig-schem}b), which confirms a direct link between the formation of spin stripes and the remarkable low-energy excitations found in this state.

\section{Discussion}

To put the observed excitations into context in terms of frustrated quantum-spin chains, we plot a schematic temperature--frequency diagram of excitations in $\beta$-TeVO$_4$ in Fig.\,\ref{fig-schem}b.
At the lowest energies, we find dielectric dynamics, which peaks at $\sim$0.4\,MHz and is most pronounced in the multiferroic VC phase (see Supplementary information \cite{sup} for complementary dielectric measurements). 
These are followed by strong $\nu_{\text{stripe}}$ fluctuations, which emerge at $T_{N2}$ and reach a maximum of 7.3(5)\,MHz close to the $T_{N3}$ transition, after which they disappear.  
The persistent spin dynamics, which can be significantly enhanced in frustrated spin-1/2 systems due to quantum effects, exhibits even faster fluctuations at 0.1--1\,GHz.
The collective magnon excitations, determined by the main exchange interactions $J_i$ ($i$\,=\,1,2), develop in the VC phase above the gap, most likely induced by spin--orbit coupling, which is also responsible for exchange anisotropy, i.e., at frequencies of 0.1--1\,THz \cite{pregelj2018coexisting, sup}.
The diagram is completed by spinon excitations, which form a continuum extending up to $\sim$$\pi J_i \sigma^2$ \cite{lake2005quantum}, in this case up to $\sim$3\,THz \cite{pregelj2018coexisting}.

\begin{figure*}[!]
\centering
\includegraphics[width=18cm]{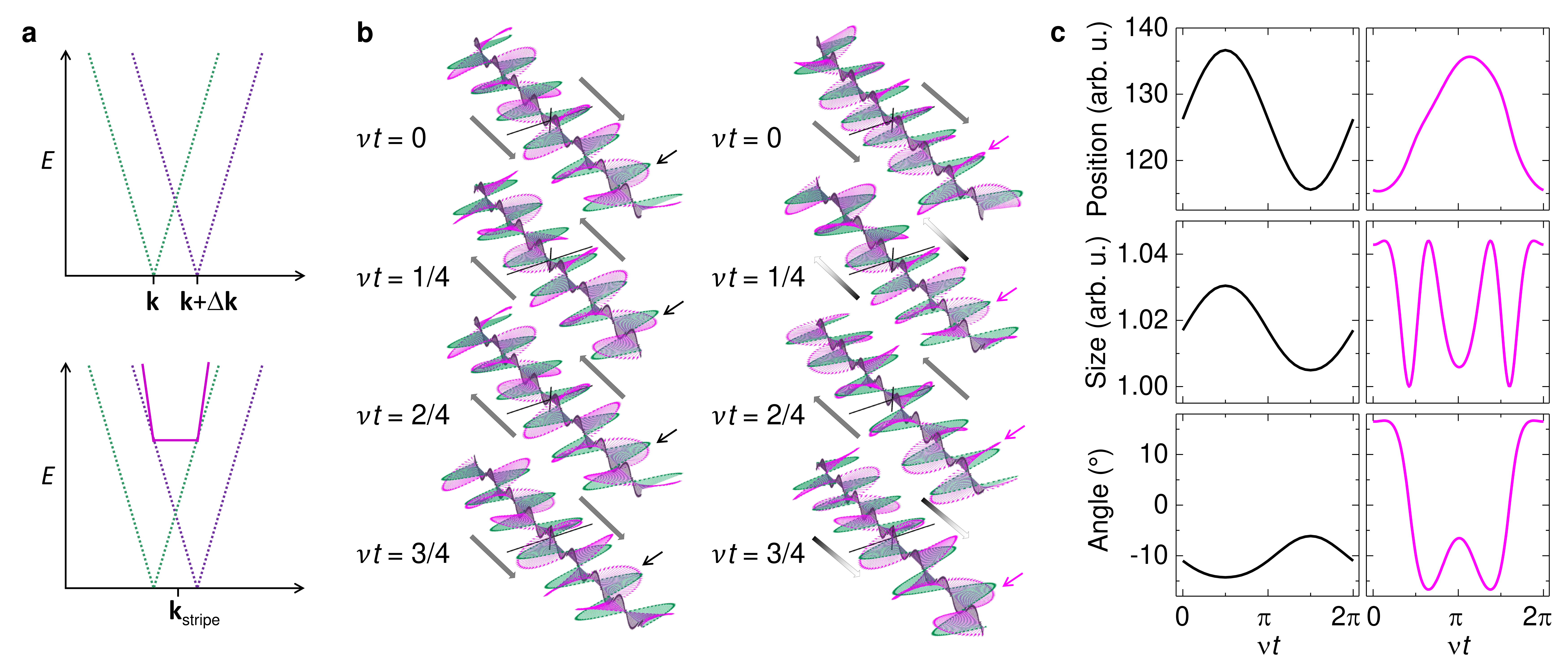}
\caption{{\bf Wigglon excitation.} {\bf a} Schematic dispersion relations for individual gapless phasons corresponding to $\sigma_{\bf k}$ and $\sigma_{{\bf k}\,+\,{\bf\Delta k}}$ (dotted lines), as well as for the gaped two-phason bound wigglon excitation (solid line).
{\bf b} Schematic representation of the magnetic order in the spin-stripe phase, decomposed into two orthogonal magnetic components (green and violet arrows) and summed together (magenta arrows), for the two decoupled phason modes (left) and for the two-phason bound wigglon excitation (right).
Consecutive figures represent the time evolution, while large arrows denote shifts of the central maximum (see simulations in Supplementary videos 1 and 2).
Note the uniform shifts for decoupled phasons and nonuniform extensions/contractions induced by a wigglon.
{\bf c} The position, amplitude and orientation of the marked maximum in the magnetic structure (thick black and magneta arrows in {\bf b}) as a function of time for decoupled phason excitations (left) and a wigglon excitation (right).}
\label{fig-wigg}
\end{figure*}

Next, we try to identify the physical mechanism responsible for the unusual spin-stripe excitations. 
Among numerous experimental and theoretical studies of spin chains, considering different $J_1/J_2$ ratios in an external magnetic field \cite{sudan2009emergent, hikihara2008vector, katsura2008quantum} as well as in the presence of magnetic anisotropy \cite{nersesyan1998incommensurate, furukawa2010chiral, huvonen2009magnetic} and interchain interactions \cite{sato2013spin, nishimoto2015interplay, weichselbaum2011incommensurate}, there appears to be no record of multi-{\bf k} magnetic structures that would resemble the spin-stripe order observed in $\beta$-TeVO$_4$, nor its associated excitations. 
Moreover, if persistent spin dynamics or any other low-energy excitation inherent to the SDW (or VC) phase, e.g., phasons \cite{blanco2013phasons, furukawa2010chiral}, were also primarily responsible for the spin relaxation in the spin-stripe phase, there should be no significant difference between the SDW (or VC) and spin-stripe dynamics.
Contrary to this, spin dynamics in the spin-stripe phase is completely different from the other ordered phases, as evidenced by the drastically enhanced muon-spin relaxation rate (Fig.\,\ref{fig-musr}). 
Further comparison with dynamical processes in spin systems that do develop multi-{\bf k} magnetic structures, e.g., skyrmion phases \cite{han2017skyrmions}, does not reveal any similarity either.
Namely, in contrast to our case, in such systems spin dynamics are typically driven either by very slow domain fluctuations in the range between 1\,Hz to 1\,kHz \cite{zang2011dynamics}, or stem from much faster collective breathing, magnon or even electromagnon excitations in the GHz range \cite{nagaosa2013topological, pimenov2006possible, mochizuki2015dynamical}. 
Finally, dynamics in alternating patterns of spin and charge stripes in oxides was found between 1 and 100\,GHz \cite{lancaster2014stripe}.
To summarize, there exists no report of electron spin dynamics in the mid-frequency (MHz) range, as observed in $\beta$-TeVO$_4$.
Such dynamics, therefore, seems to originate from the peculiarities of the spin-stripe phase, which are also responsible for the remarkable coincidence of the $\nu_{\text{stripe}}(T)$ and $F_4(T)$ dependences (Fig.\,\ref{fig-schem}b).

To explain the sequence of the magnetic transitions as well as to clarify the existence of the spin-stripe phase and its corresponding excitations we undertake a phenomenological approach based on the classical Ginzburg--Landau theory of phase transitions.
We construct the following expression for the magnetic free energy
\begin{eqnarray}
\label{GLmodel2}
F & = & A_1 \left(T/T_{N1} - 1\right) \sigma_{\bf k}^2 + A_2 \left(T/T_{N2} - 1\right) \sigma_{{\bf k+\Delta k}}^2 +\nonumber\\
& & B_1 \sigma_{\bf k}^4 + B_2 \sigma_{{\bf k+\Delta k}}^4 + C \left(1 - \Delta q\right)^2 \sigma_{{\bf k+\Delta k}}^2 +F_4,\\
F_4 & = & \left[D + E f(\sigma_{\bf k},\sigma_{{\bf k+\Delta k}}) \Delta q^2\right] \sigma_{\bf -k}^2 \sigma_{{\bf k+\Delta k}} \sigma_{{\bf k-\Delta k}},
\end{eqnarray}
where $A_i$, $B_i$ ($i$\,=\,1,2), $C$, $D$, and $E$ are scaling constants.
The first four terms in (\ref{GLmodel2}) describe the evolution of two independent magnetic order parameters $\sigma_{\bf k}$ and $\sigma_{{\bf k+\Delta k}}$ that emerge at $T_{N1}$ and $T_{N2}$, respectively.
The fifth term represents exchange anisotropy that is responsible for a different magnetic wave vector for the $\sigma_{{\bf k+\Delta k}}$ component, i.e., favoring $\Delta q$\,=\,$\Delta k/\Delta k_0$\,$\neq$\,0, where $\Delta k_0$\,$\equiv$\,$\Delta k(T_{N2})$ represents the discrepancy between the native magnetic wave vectors for the $\sigma_{{\bf k+\Delta k}}$ and $\sigma_{\bf k}$ components.
The $F_4$ term is associated with the coupling between the two order parameters and favors fully developed magnetic moments, i.e., it acts against the discrepancy between the two modulation periods ($\Delta q$\,$\to$\,0).
The function $f(\sigma_{\bf k},\sigma_{{\bf k+\Delta k}})$ accounts for the size limitation of the V$^{4+}$ $S$\,=\,1/2 magnetic moments and thus smoothly changes from 0 to 1, when $\sqrt{\sigma_{\bf k}^2 + \sigma_{{\bf k+\Delta k}}^2}$ exceeds the limiting value \cite{sup}. 
Considering $\sigma_{{\bf k-\Delta k}}$\,$\approx$\,$\sqrt{0.2} \sigma_{{\bf k+\Delta k}}$  (Fig.\,\ref{fig-NDphen} and Ref.\,\onlinecite{pregelj2015spin}), the minimization of (\ref{GLmodel2}) with respect to $\Delta q$, $\sigma_{\bf k}$ and $\sigma_{{\bf k+\Delta k}}$ \cite{sup} returns the corresponding temperature dependences (Fig.\,\ref{fig-NDphen}b and \ref{fig-NDphen}c) that almost perfectly describe the observed behavior.
In particular, we find that in the vicinity of the paramagnetic phase, where ordered magnetic moments are still small, a sizable exchange anisotropy can impose different modulations for different magnetic-moment components through the $C$ term.
On cooling, however, the ordered magnetic moments increase, causing the $F_4$ term to prevail and thus to stabilize the VC phase with $\Delta q$\,=\,0 below $T_{N3}$.
Finally, the derived parameters allow us to calculate the temperature dependence of the $F_4$ term. 
Comparison of the derived temperature dependence $F_4(T)$ with experimentally determined $\nu_{\text{stripe}}(T)$, i.e., assuming that $h \nu_{\text{stripe}}$\,=\,$c|F_4|$, where $h$ is the Planck constant, we obtain a very good agreement for $c$\,=\,0.7 (Fig.\,\ref{fig-schem}b), corroborating the connection between the $F_4$ term and the $\nu_{\text{stripe}}$ dynamics.

Having established the intimate relation between low-frequency excitations and the spin-stripe phase, the open question that remains concerns the microscopic nature of the spin-stripe excitation mode.
In contrast to ordinary magnon and spinon modes, these excitations arise from a fourth-order free-energy term that couples magnetic components with different modulation periods (Fig.\,\ref{fig-schem}a).
Higher-order terms in the free energy impose an interaction between the basic elementary excitations \cite{torrance1969excitation, kecke2007multimagnon, nawa2017dynamics}. 
The observed excitations are thus most likely bound states of two elementary excitations of the incommensurate amplitude-modulated magnetic components. 
The latter may either be two phasons, i.e., linearly dispersing zero-frequency Goldstone modes that change the phase of the modulation \cite{blanco2013phasons, overhauser1971observability}, two amplitudons, i.e., high-frequency modes that change the amplitude of the modulation \cite{blanco2013phasons}, or a combination of the two.
Given that $\nu_{\text{stripe}}$ is very small compared to the exchange interactions, the spin-stripe excitation is most likely a two-phason bound mode that has minimal energy at a certain wave vector {\bf k}$_{\text{stripe}}$ (Fig.\ref{fig-wigg}a).
Since {\bf k}$_{\text{stripe}}$, in principle, differs from both {\bf k} and {\bf k}$\pm${\bf$\Delta$k}, the bound mode imposes additional expansion and contraction of the modulation periods on top of the phase changes induced by individual phasons (Fig.\,\ref{fig-wigg}b).
Consequently, positions, sizes and orientations of maxima in the magnetic structure exhibit completely different time dependences than for individual phason (Fig.\,\ref{fig-wigg}c), resulting in a wiggling-like motion of the magnetic moments (see simulation in Ref.\,\onlinecite{sup}).
Hence, we dubbed this type of spin-stripe excitations ``wigglons''.
Moreover, the corresponding amplitude variation is reminiscent of the longitudinal (amplitude) Higgs mode in the Ca$_2$RuO$_4$ antiferromagnet, which has also been found to decay into a pair of Goldstone modes \cite{jain2017higgs}.
Finally, we point out that ``wigglon'' dynamics in the spin-stripe phase of $\beta$-TeVO$_4$ might share similarities with fluctuating-charge-stripe phases \cite{klauss2012musr, kivelson2003detect}, where the nematic response is ascribed to fast stripe dynamics \cite{anissimova2014direct}.

Our results reveal an intriguing spin-only manifestation of fluctuating-stripe physics, which has, so far, been studied exclusively in the context of nematic phases in high-temperature superconductors.
We show that $\beta$-TeVO$_4$ displays an intriguing spin-stripe order, which, due to a slow wiggling motion of the magnetic moments, appears static on the neutron-scattering timescale \cite{pregelj2015spin}, i.e., at $\nu$\,$>$\,10\,GHz, while it is in fact dynamical at MHz frequencies.
The phenomenon is driven by sizable exchange anisotropy, which prevails in a finite temperature range where it stabilizes the dynamical spin-stripe phase that hosts an extraordinary type of excitation, driven by the fourth-order coupling term in the magnetic free energy.
Our discovery draws attention to other frustrated spin-1/2 chain compounds with complicated and unresolved magnetic phase diagrams \cite{willenberg2016complex, bush2018exotic}, where similar effects may be anticipated to play a role.
Finally, more details of the wigglon excitation, such as their dependence on the applied magnetic field, should be explored by complementary nuclear-magnetic-resonance measurements that are highly sensitive to spin dynamics in the relevant MHz range even in a sizable applied magnetic field.

\section{Methods}

{\bf Sample description.} 
The single-crystal samples were grown from TeO$_2$ and VO$_2$ powders by chemical vapor transport reaction, using two-zone furnace and TeCl$_4$ as a transport agent, as explained in Ref.\,\onlinecite{pregelj2015spin}.
Powder samples were obtained by grinding single-crystal samples.

{\bf $\mu$SR experiments.}
The experiments were performed on the MuSR instrument at the STFC ISIS facility, Rutherford Appleton Laboratory, United Kingdom, and on the General Purpose Surface-Muon instrument (GPS) at
the Paul Scherrer Institute (PSI), Switzerland.
The measurements at PSI were performed in zero-field, while at ISIS a small longitudinal field of 4\,mT was applied to decouple the muon relaxation due to nuclear magnetism at long times.
The dead time at the GPS instrument is $\sim$0.01$\mu$s, whereas its is $\sim$0.1\,$\mu$s at the MuSR instrument.
For details on background subtraction see Ref.\,\onlinecite{sup}.

{\bf Neutron diffraction.}
Neutron diffraction measurements were performed on a 2$\times$3$\times$4\,mm$^3$ single-crystal on the triple-axis-spectrometer TASP at PSI.
To assure the maximal neutron flux the wavelength of 3.19\,\AA\, was chosen for the experiment.
An analyzer was used to reduce the background, while the standard ILL orange cryostat was used for cooling.

\section{Data availability}
The data that support the findings of this study are available from the corresponding author upon reasonable request.

\section{Acknowledgments}

\begin{acknowledgments}

We thank T. Lancaster for fruitful discussions and valuable comments.
We are grateful for the provision of beam time at the Science and Technology Facilities Council (STFC) ISIS Facility, Rutherford Appleton Laboratory, UK, and S$\mu$S, Paul Scherrer Institut, Switzerland.
This work has been funded by the Slovenian Research Agency (project J1-9145 and program No. P1-0125), the Swiss National Science Foundation (project SCOPES IZ73Z0\_152734/1) and the Croatian Science Foundation (project IP-2013-11-1011).
This research project has been supported by the European Commission under the 7th Framework Programme through the 'Research Infrastructures' action of the 'Capacities' Programme, NMI3-II Grant number 283883, Contract No. 283883-NMI3-II.
M.G. is grateful to EPSRC (UK) for financial support (grant No. EP/N024028/1).
We are grateful to M. Enderle for local support at Institut Laue-Langevin, Grenoble, France.

\end{acknowledgments}

\section{Competing interests}
 The authors declare no competing interests.

\section{Author contributions}
M.P., A.Z., and D.A. designed and supervised the project.
The samples were synthesized by H.B. 
The $\mu$SR experiments were performed by A.Z., M.G., H.L. and F.C and analyzed by M.P., M.G., and A.Z..
The neutron diffraction experiments were performed by M.P., O.Z. and J.S.W. and analyzed by M.P. 
The dielectric experiments were performed by T.I. and D.R.G.
All authors contributed to the interpretation of the data and to the writing of the manuscript.


%

\end{document}